# 3D Analytical Model of Skyrmions and Skyrmion-like Structures in a Two-sublattice Antiferromagnet with Dzyaloshinskii-Moriya Interaction

O.Yu. Gorobets[1,2*], Yu.I. Gorobets[1,2], V.S. Kovalenko[1]

[1]*National Technical University of Ukraine «Igor Sikorsky Kyiv Polytechnic Institute», 37 Peremohy Ave., 03056 Kyiv, Ukraine*

[2]*Institute of Magnetism of NAS and MES of Ukraine, 36b Acad. Vernadskoho Blvd., 03142 Kyiv, Ukraine*

*Correspondence to: Email: gorobets.oksana@gmail.com

Today antiferromagnetic spintronics is a rapidly developing new field of research of antiferromagnetic materials as elements in spintronic devices. Antiferromagnets represent perspective materials for spintronic generators where THz-frequencies can be reached enabling ultrafast information processing in comparison with ferromagnets [1-5]. However, antiferromagnets have not been attractive for spintronics for a long time because they have compensated magnetization of the ground state due to existence of several magnetic sublattices with strong exchange coupling, and it was hard enough to detect magnetization and staggered magnetization (i.e. Neel order vector) of antiferromagnet as well as it is necessary to apply strong magnetic field to manipulate both order parameters. Recently these difficulties have been solved. Electrical detection of antiferromagnetic ordering was realized on basis of (tunneling) anisotropic magnetoresistance effect [6-9]. There is also detection based on spin pumping effect when motion of Neel order vector induces spin pumping effect of a similar magnitude as in ferromagnets [10, 11]. There are two methods of manipulation antiferromagnetic ordering using electric [12, 13] and spin currents [14-17]. Recently, electric manipulation of antiferromagnetic Neel vector was predicted theoretically [12] and confirmed experimentally [13]. It is based on the effect that electric current in antiferromagnets can induce Neel order field through spin-orbit interaction, which is different for different magnetic sublattices [12, 13].

There are two most popular candidates as information carries for the next generation memory devices in antiferromagnets – antiferromagnetic domain walls and antiferromagnetic skyrmions. The skyrmions represent stable nanoscale magnetization nonuniformities in ferro- and antiferromagnets [18, 19]. Usually the skyrmions are topologically protected and the stability of skyrmions is supplied by the Dzyaloshinskii-Moriya interaction [20, 21], which is commonly found not only for ferromagnets, but for in antiferromagnets as well [22, 23].

In this paper, the analytical model is developed for description of skyrmions and skyrmion-like magnetic structures in a two-sublattice antiferromagnet with uniaxial magnetic anisotropy and Dzyaloshinskii-Moriya interaction. "Relativistic contraction" of skyrmion size in the direction of motion is demonstrated for "subcritical" case when the skyrmion velocity is less than spin wave velocity in antiferromagnet. Lorentz-like "supercritical" transformation are found for skyrmion-like magnetic structures moving with velocity greater than spin wave velocity in antiferromagnet.

Let us consider a two-sublattice antiferromagnet with uniaxial magnetic anisotropy, Dzyaloshinskii-Moriya interaction and magnetizations of sublattices $\vec{M}_1, \vec{M}_2$, $\left|\vec{M}_1\right| = \left|\vec{M}_2\right| = M_0$ where absolute value of magnetizations of both sublattices are equal to $M_0 = const$.

The Landau-Lifshitz equations for a two-sublattice antiferromagnet have the form according to [24, 25]:

$$\begin{cases} \dfrac{2}{gM_0}\dfrac{\partial \vec{m}}{\partial t} = \left[\vec{m}\times H_m^{(ef)}\right]+\left[\vec{l}\times H_l^{(ef)}\right], & H_m^{(ef)} = -\dfrac{1}{M_0^2}\dfrac{\delta W}{\delta \vec{m}} \\ \dfrac{2}{gM_0}\dfrac{\partial \vec{l}}{\partial t} = \left[\vec{l}\times H_m^{(ef)}\right]+\left[\vec{m}\times H_l^{(ef)}\right], & H_l^{(ef)} = -\dfrac{1}{M_0^2}\dfrac{\delta W}{\delta \vec{l}} \end{cases} \quad (1)$$

$g = \dfrac{2\mu_0}{\hbar}$, $\mu_0$ is Bohr magneton, $\hbar$ is Planck constant, $W$ is an energy of an antiferromagnet. The dimensionless vectors of antiferromagnetism $\vec{l}$ and magnetization $\vec{m}$ are introduced in (1):

$$\vec{l} = \frac{\vec{M}_1 - \vec{M}_2}{2M_0}, \quad \vec{m} = \frac{\vec{M}_1 + \vec{M}_2}{2M_0}, \quad (2)$$

where $\vec{M}_1$, $\vec{M}_2$ are magnetizations of sublattices.

Due to the definition, it is possible to write down the constraint
$$\vec{m}\vec{l} = 0, \quad \vec{m}^2 + \vec{l}^{\,2} = 1. \quad (3)$$

If $\vec{m}^2 \ll \vec{l}^{\,2}$ than
$$\vec{l}^{\,2} \approx 1. \quad (4)$$

The equations (1) describe the excited states of an antiferromagnet. The magnetic energy of a uniaxial antiferromagnet with Dzyaloshinskii-Moriya interaction has the form [24, 25]:

$$W = M_0^2 \int dV \left\{ \frac{\delta}{2}\vec{m}^2 + \frac{\alpha}{2}(\nabla \vec{l})^2 + \frac{\alpha'}{2}(\nabla \vec{m})^2 + \frac{\beta_1}{2}l_z^2 + \frac{\beta_2}{2}m_z^2 + \vec{d}\left[\vec{m}\times\vec{l}\right] - \vec{m}\vec{h}_0 \right\}, \quad (5)$$

where $\delta$ is uniform exchange constant, $\alpha$, $\alpha'$ are the nonuniform exchange constants ($\alpha > 0$), $\beta_1$, $\beta_2$ are the constants of uniaxial magnetic anisotropy, $\vec{d}$ is Dzyaloshinskii-Moriya vector, directed parallel to the principal axis of the crystal, $\vec{h}_0 = \dfrac{2\vec{H}_0}{M_0}$, $\vec{H}_0$ is an external magnetic field, integration in (5) is taken over the volume of an antiferromagnet. If $\beta_1 < 0$ than the magnetic moments of sublattices are directed along the anisotropy axis at the ground state (an antiferromagnet with an easy axis magnetic anisotropy), and if $\beta_1 > 0$ than the minimum of energy is reached for vector $\vec{l}$ perpendicular to the anisotropy axis (an antiferromagnet with an easy plane magnetic anisotropy). The ground state is characterized by $\vec{m} = 0$ if $\beta_1 < -\dfrac{d^2}{2\delta} + \dfrac{h_0^2}{2\delta^2}(\beta_2 - \delta)$ and $\vec{m} \neq 0$ otherwise due to DMI.

The terms with gradients of magnetization can be neglected in the second equation of the system (1) under long wavelength approximation. In this case, this equation can be approximately written as follows:

$$\frac{2}{gM_0}\frac{\partial \vec{l}}{\partial t} = \left[\delta \vec{m}\times\vec{l}\right] + \left[\vec{l}\times\left[\vec{d}\times\vec{l}\right]\right] + \left[\vec{l}\times\vec{h}_0\right] \quad (6)$$

If $\vec{m}^2 \ll \vec{l}^{\,2}$ and $\beta \ll d \ll \delta$ [24] than the magnetization $\vec{m}$ can be derived from the last equation:

$$\vec{m} = \frac{2}{\delta g M_0}\left[\vec{l}\times\frac{\partial \vec{l}}{\partial t}\right] + \frac{1}{\delta}\left[\vec{d}\times\vec{l}\right] - \frac{1}{\delta}\vec{l}\left(\vec{h}_0\cdot\vec{l}\right) + \frac{1}{\delta}\vec{h}_0. \quad (7)$$

The condition is used in (7):

$$\vec{m}^2 << \vec{l}^{\,2} \square 1. \quad (8)$$

The relation (7) is valid under not very strong magnetic fields $H_0 << \delta M_0$ and not very high frequencies $\hbar\omega << \mu_0 \delta M_0$.

The spatial derivatives of vector $\vec{m}$ can be neglected in the first equation of the system (1) under the same approximation. Substituting (7), one can obtain the equation of dynamics of antiferromagnetic vector $\vec{l}$ in a two-sublattice antiferromagnet with Dzyaloshinskii-Moriya interaction:

$$\left[\vec{l} \times \left(\alpha\Delta\vec{l} - \frac{4}{\delta(gM_0)^2}\frac{\partial^2 \vec{l}}{\partial t^2}\right)\right] + \frac{4}{\delta gM_0}\frac{\partial \vec{l}}{\partial t}(\vec{l}\cdot\vec{h}_0) - \frac{2}{\delta gM_0}\frac{\partial \vec{h}_0}{\partial t}$$
$$-\beta_1(\vec{l}\cdot\vec{e}_z)[\vec{l}\times\vec{e}_z] + \frac{1}{\delta}(\vec{d}\cdot\vec{l})[\vec{d}\times\vec{l}] + \frac{1}{\delta}[\vec{l}\times[\vec{h}_0\times\vec{d}]] - \frac{1}{\delta}[\vec{l}\times\vec{h}_0](\vec{l}\cdot\vec{h}_0) = 0 \quad (10)$$

The equation of dynamics of antiferromagnetic vector (10) can be written though the angular variables that are introduces by the standard way:

$$l_x = \sin\theta\cos\varphi, \; l_y = \sin\theta\sin\varphi, \; l_z = \cos\theta, \quad (11)$$

where $\theta$ and $\varphi$ are the polar and azimuth angles for the antiferromagnetic vector, $l_x$, $l_y$, $l_z$ are the Cartesian coordinates of the antiferromagnetic vector. If the Dzyaloshinskii vector and the external magnetic field are directed along OZ axis than the equation (10) has the form:

$$\begin{cases} c^2\Delta\theta - \dfrac{\partial^2\theta}{\partial t^2} - \left[c^2(\nabla\varphi)^2 - \left(\dfrac{\partial\varphi}{\partial t} - \omega_H\right)^2 - \omega_0^2 \,\text{sgn}\left(\dfrac{d^2}{\delta} + \beta_1\right)\right]\sin\theta\cos\theta = 0 \\ c^2 div(\sin^2\theta\nabla\varphi) - \dfrac{\partial}{\partial t}\left[\sin^2\theta\left(\dfrac{\partial\varphi}{\partial t} - \omega_H\right)\right] + \dfrac{\partial\omega_H}{\partial t} = 0 \end{cases}, \quad (12)$$

where $\omega_H = \dfrac{|g|M_0 h_0}{2}$, $c = \dfrac{\sqrt{\alpha\delta}|g|M_0}{2}$, $\omega_0^2 = \dfrac{c^2}{\alpha}\left|\dfrac{d^2}{\delta} + \beta_1\right|$. It is important to note that the Lorentz-invariance is present even for non-zero DMI, whereas the presence of the magnetic field destroy this invariance, but keep a special kind of Lorentz-covariance, see [26, 27] and for details recent review [28].

If an external magnetic field is constant than the system of equations (12) can be transformed to the system of equations considered in papers [29-33]. That is why using the results of investigations [29-33] it is easy to show by means of changing notations of coefficients that the system of equations (12) has the following solution on the basis of the results of the papers [29-33]. In particular, if $-\dfrac{1}{4} < C < 0$ one can obtain the solution:

$$\begin{cases} tg\left(\dfrac{\theta}{2}\right) = \dfrac{b}{dn\left(a\sqrt{|C|}P(X,Y,Z), k_1\right)}, \\ \varphi = \omega_H t + Q(X,Y,Z) \end{cases} \quad (13)$$

where

$$a = \sqrt{\dfrac{1+2C+\sqrt{1+4C}}{2|C|}}, \; b = \sqrt{\dfrac{1+2C-\sqrt{1+4C}}{2|C|}}, \; k_1 = \sqrt{\dfrac{2\sqrt{1+4C}}{1+2C+\sqrt{1+4C}}}, \quad (14)$$

and if $C>0$ than there is a solution:

$$\begin{cases} tg\left(\dfrac{\theta}{2}\right) = \sqrt{\dfrac{1 - sn\left(\dfrac{P(X,Y,Z)}{k_2}, k_2\right)}{1 + sn\left(\dfrac{P(X,Y,Z)}{k_2}, k_2\right)}}, \\ \varphi = \omega_H t + Q(X,Y,Z) \end{cases} \quad (15)$$

where $k_2 = 1/\sqrt{1+4C}$. Here and in the Eq.(14) above $sn(u,k)$ and $dn(u,k)$ are the Jacobi elliptic functions with the elliptic modulus $k$.

Let us apply the Lorentz-like transformations of coordinates at $v<c$:

$$\begin{cases} X = \dfrac{x-vt}{l_0\sqrt{1-\dfrac{v^2}{c^2}}} \\ Y = \dfrac{y}{l_0} \\ Z = \dfrac{z}{l_0} \end{cases} \text{ or } \begin{cases} X = \dfrac{x}{l_0} \\ Y = \dfrac{y-vt}{l_0\sqrt{1-\dfrac{v^2}{c^2}}} \\ Z = \dfrac{z}{l_0} \end{cases} \text{ or } \begin{cases} X = \dfrac{x}{l_0} \\ Y = \dfrac{y}{l_0} \\ Z = \dfrac{z-vt}{l_0\sqrt{1-\dfrac{v^2}{c^2}}} \end{cases}. \quad (16)$$

The Lorentz-like transformations of coordinates (16) at subcritical self-similar motion at $v<c$ mean that the equations (12) are Lorentz invariant and any nonlinear solution of these equations can move straightly with constant velocity along any coordinate axis.

If $v>c$ than one can apply the supercritical Lorentz-like transformations of coordinates instead of (16):

$$\begin{cases} X = \dfrac{x}{l_0} \\ Y = \dfrac{y}{l_0} \\ Z = \dfrac{z-vt}{l_0\sqrt{\dfrac{v^2}{c^2}-1}} \end{cases}. \quad (17)$$

$$l_0 = \dfrac{c}{\omega_0} = \sqrt{\dfrac{\alpha}{\left|\dfrac{d^2}{\delta}+\beta_1\right|}}$$

The Lorentz-like supercritical transformations of coordinates (17) at supercritical self-similar motion at $v>c$ do not follow directly from the equations (12) because the equations (12) are not invariant relative to the supercritical Lorentz-like transformations (17). However, it will be shown further that there are specific solutions of the equations (12) which can move straightly with constant velocity $v>c$ along only one coordinate axis (it is OZ axis according to the notations chosen in this paper).

The functions $P(X,Y,Z)$ and $Q(X,Y,Z)$ can be chosen in the form:

$$P(X,Y,Z) = Z\Theta\left[-\left(\frac{d^2}{\delta} + \beta_1\right)\cdot\left(1 - \frac{v^2}{c^2}\right)\right] + f(X,Y)$$

$$Q(X,Y,Z) = Z\Theta\left[\left(\frac{d^2}{\delta} + \beta_1\right)\cdot\left(1 - \frac{v^2}{c^2}\right)\right] + g(X,Y)$$
(18)

where the conjugated harmonic functions $f(X,Y)$ and $g(X,Y)$ have the form in the solution (15):

$$\begin{cases} f(X,Y) = \sum_i \tilde{\tilde{n}}_i \ln\left(|\vec{r} - \vec{r}_{0i}|\right) + \frac{2}{\pi} k_2 \cdot K(k_2) \sum_i \tilde{n}_i \alpha_i + C_2 + \\ + \sum_i \sum_n \frac{A_n^{(i)}}{|\vec{r} - \vec{r}_{0i}|^n} \left(B_n^{(i)} \cos n\alpha_i + C_n^{(i)} \sin n\alpha_i\right) \\ g(X,Y) = -\frac{2}{\pi} k_2 \cdot K(k_2) \sum_i \tilde{n}_i \ln\left(|\vec{r} - \vec{r}_{0i}|\right) + \sum_i \alpha_i \tilde{\tilde{n}}_i + C_3 + \\ + \sum_i \sum_n \frac{A_n^{(i)}}{|\vec{r} - \vec{r}_{0i}|^n} \left(C_n^{(i)} \cos n\alpha_i - B_n^{(i)} \sin n\alpha_i\right) \end{cases}$$
(19)

Here the following notations are introduced: $\vec{r}$ is the two dimensional vector with coordinates in a plane $XOY$ $\vec{r} = (X,Y)$, $\vec{r}_{0i}$ is the two dimensional vector with coordinates in a plane $YOZ$ which is perpendicular to the direction of propagation of a nonlinear spin wave, $\vec{r}_{oi} = (X_{0i}, Y_{0i})$, $X_{0i}$, $Y_{0i}$ are the dimensionless constants, $\tan\alpha_i = \frac{Y - Y_{0i}}{X - X_{0i}}$, $i$, $n$, $\tilde{n}_i$, $\tilde{\tilde{n}}_i$ are the integer numbers, $\Theta(\xi)$ is the Heaviside step function and $K(k)$ is complete elliptic integral of the first kind

$$\Theta(\xi) = \begin{cases} 0, & \xi \leq 0 \\ 1, & \xi > 0 \end{cases}, \quad K(k) = \int_0^{\frac{\pi}{2}} \frac{d\xi}{\sqrt{1 - k^2 \sin^2 \xi}}.$$
(20)

Let's note that the function $f(X,Y)$ in (19) represents the expansion into series of powers of $|\vec{r} - \vec{r}_{0i}|$ of an arbitrary harmonic function of two variables $X$ and $Y$, and the expression for the function $g(X,Y)$ in (19) represents the expansion into series of powers of $|\vec{r} - \vec{r}_{0i}|$ of harmonic function of two variables $X$ and $Y$ which is conjugated to the function $f(X,Y)$. It means that the functions $f(X,Y)$ and $g(X,Y)$ are connected by the Cauchy-Riemann conditions and are the Eigen functions of the two dimension Laplace operator.

The formula (7) allows calculating the magnetization of an antiferromagnet:

$$\begin{cases} m_x = \dfrac{h_0}{\omega_H \delta}\left\{\sin\varphi\dfrac{\partial\theta}{\partial t} + \sin\theta\cos\theta\cos\varphi\dfrac{\partial\varphi}{\partial t}\right\} - \\ \quad -\dfrac{d}{\delta}\sin\theta\sin\varphi - \dfrac{h_0}{\delta}\sin\theta\cos\theta\cos\varphi \\ m_y = -\dfrac{h_0}{\omega_H \delta}\left\{\cos\varphi\dfrac{\partial\theta}{\partial t} - \sin\theta\cos\theta\sin\varphi\dfrac{\partial\varphi}{\partial t}\right\} + \\ \quad +\dfrac{d}{\delta}\sin\theta\cos\varphi - \dfrac{h_0}{\delta}\sin\theta\cos\theta\sin\varphi \\ m_z = \dfrac{h_0}{\delta}\sin^2\theta\left(1 - \dfrac{1}{\omega_H}\dfrac{\partial\varphi}{\partial t}\right) \end{cases} \qquad (21)$$

It is easy to show that the general form of the solution (15)-(19) includes the particular solution

$$\cos\theta = sn\left(\dfrac{Z + \ln\dfrac{|\vec{r}|}{R_1}}{k_2}, k_2\right), \quad \varphi = \alpha + \alpha_0 \qquad (22)$$

for an easy axis antiferromagnet when $\left(\dfrac{d^2}{\delta} + \beta_1\right) < 0$ and $v < c$. The solution (22) is valid also when $\left(\dfrac{d^2}{\delta} + \beta_1\right) > 0$ and $v > c$. $\alpha_0$ is an arbitrary constant in (22). The solution (22) describes, for example, an antiferromagnetic domain wall with skyrmion-like structure in a long cylindrical nano-shell with inner radius $R_1$ and outer radius $R_2 = R_1 \exp(2k_2 K(k_2))$.

If the modulus of an elliptic function $k_2 = 1$, the solution (22) can be transformed to the solution with an arbitrary characteristic scale $R$

$$\cos\theta = th\left(Z + \ln\dfrac{|\vec{r}|}{R}\right), \quad \varphi = \alpha + \alpha_0. \qquad (23)$$

The solution (23) describes both antiferromagnetic skyrmion with a characteristic size $l_0$ in a thin antiferromagnetic plate with thickness much less than the skyrmion size $l_0$ (in this case, it is possible to neglect dependence on $Z$ coordinate in (23)) and an antiferromagnetic domain wall with skyrmion-like structure in a long antiferromagnetic nanowire. The solution of type (23) was considered in [34] for description of vortice state of an antiferromagnet without Dzyaloshinskii-Moriya interaction.

Fig. 1 a) represents $l_z = \cos\theta$ as a function of $r = |\vec{r}|$ at $Z = 0$ according to the formula (23).

The skyrmions and antiferromagnetic domain walls with skyrmion-like structure (22), (23) can move straightly over long distance without the skyrmion Hall effect (i.e. with zero Magnus force) similarly to the skyrmions in antiferromagnets in the papers [23, 26, 27, 35-37]. The skyrmions (23) demonstrate Lorentz-like length contraction in the direction of movement at $v < c$ because according to (16) the characteristic size of a skyrmion is $l_0$ at $v = 0$ and it is $l = l_0\sqrt{1 - \dfrac{v^2}{c^2}}$ for a skyrmion moving with velocity $v < c$. The Lorentz like length contraction leads to deformation of skyrmion in the case of its movement in plane of an antiferromagnetic plate (i.e. along OX or OY axes). Similar stretching of an antiferromagnetic skyrmion in the direction perpendicular to its velocity was observed in micromagnetic simulation in paper [37]. However, the concrete velocity dependence of skyrmion shape was not investigated in [37]. If $\alpha_0 = \frac{\pi}{2}$ than the solution (23) describes Neel antiferromagnetic skyrmion. If $\alpha_0 = 0$ than the solution (23) describes Bloch antiferromagnetic skyrmion.

In the case of movement of antiferromagnetic domain wall with skyrmion-like structure along the symmetry axis of long nano-wire or nano-shell (i.e. OZ direction), the Lorentz like length contraction does not change the round shape of skyrmion-like structure in plane XOY according to (22) and (23).

The solutions (22) and (23) demonstrate Lorentz-like supercritical length deformation of an antiferromagnetic domain wall with skyrmion-like structure moving in along the symmetry axis of a long wire or nanoshell (i.e. in the direction of movement along OZ axis) at $v > c$. According to (17) the characteristic size of an antiferromagnetic domain wall with skyrmion-like structure is $l = l_0 \sqrt{\frac{v^2}{c^2} - 1}$ for the motion with velocity $v > c$. As the supercritical movement is possible only along OZ direction, the Lorentz like supercritical length deformation does not change the round shape of skyrmion-like structure of antiferromagnetic domain wall in plane XOY. There is no the same type of skyrmion-like solution at $v = 0$ as at $v > c$ in the case of supercritical movement. The size of an antiferromagnetic domain wall is approaching to zero in the direction OZ at $v > c$ when $v$ is approaching $c$. The simple analysis of the formula (17) shows that the size of an antiferromagnetic domain wall with skyrmion-like structure along OZ direction is less than its size $l_0$ in plane XOY at $c < v < \sqrt{2}c$. The size of an antiferromagnetic domain wall with skyrmion-like structure along OZ direction is greater than its size $l_0$ in plane XOY at $v > \sqrt{2}c$. And size of an antiferromagnetic domain wall with skyrmion-like structure along OZ direction is equal to its size $l_0$ in plane XOY at $v = \sqrt{2}c$.

The solution (23) corresponds to magnetization in a cylindrical coordinate system if an external magnetic field $h_0 = 0$

$$m_\rho = 0, \quad m_\alpha = \pm \frac{d}{\delta} \sqrt{1 - th^2\left(Z + \ln\frac{|\vec{r}|}{R}\right)}, \quad m_z = 0. \qquad (24)$$

The magnetization distribution (24) does not create a magnetostatic field so the last is zero $\vec{H}^{(m)} = 0$ at $v = 0$. Fig. 1 b) represents $\frac{\delta m_\alpha}{d}$ as a function of $r = \frac{|\vec{r}|}{R}$ at $Z = 0$ according to the formula (24). The magnetizations of magnetic sublattices of an antiferromagnet are compensated in the ground state (i.e. zero ground state magnetization of an antiferromagnet) and the antiferromagnetic skyrmion size is defined by the diameter of the circle, where z-components of magnetizations of both magnetic sublattices is zero [36, 37]. The solution (23) describes the antiferromagnetic skyrmions in an antiferromagnet with zero magnetization of the ground state like the results of the papers [36, 37]. Fig. 1. represents $l_z = \cos\theta$ as a function of $r = \frac{|\vec{r}|}{R}$ at $Z = 0$ according to the formula (23) and $\frac{\delta m_\alpha}{d}$ as a function of $r = \frac{|\vec{r}|}{R}$ at $Z = 0$ according to the formula (24). The distance from the center of a skyrmion is represented in $l_0$ units in Fig. 1, Fig. 2 and Fig. 3. The material parameter ranges have been recalculated on basis of parameters of the papers [37]: $\delta \approx 100$, $d \approx (0-60)$, $\alpha \approx 0.3 \cdot 10^{-12}$ cm², $\beta_1 \approx (0-10)$. If $\delta \approx 100$, $d \approx 60$, $\alpha \approx 0.3 \cdot 10^{-12}$ cm², $\beta_1 \approx 0$ the length $l_0 \approx 7$ nm and $l_0$ increases if $d$ decreases. For example, if $\delta \approx 100$, $d \approx 3$, $\alpha \approx 0.3 \cdot 10^{-12}$ cm², $\beta_1 \approx 0$ than $l_0 \approx 30$ nm. But if the uniaxial magnetic anisotropy parameter isn't very small, for example, $\delta \approx 100$, $d \approx 60$, $\alpha \approx 0.3 \cdot 10^{-12}$ cm², $\beta_1 \approx 10$ than $l_0 \approx 1$ nm

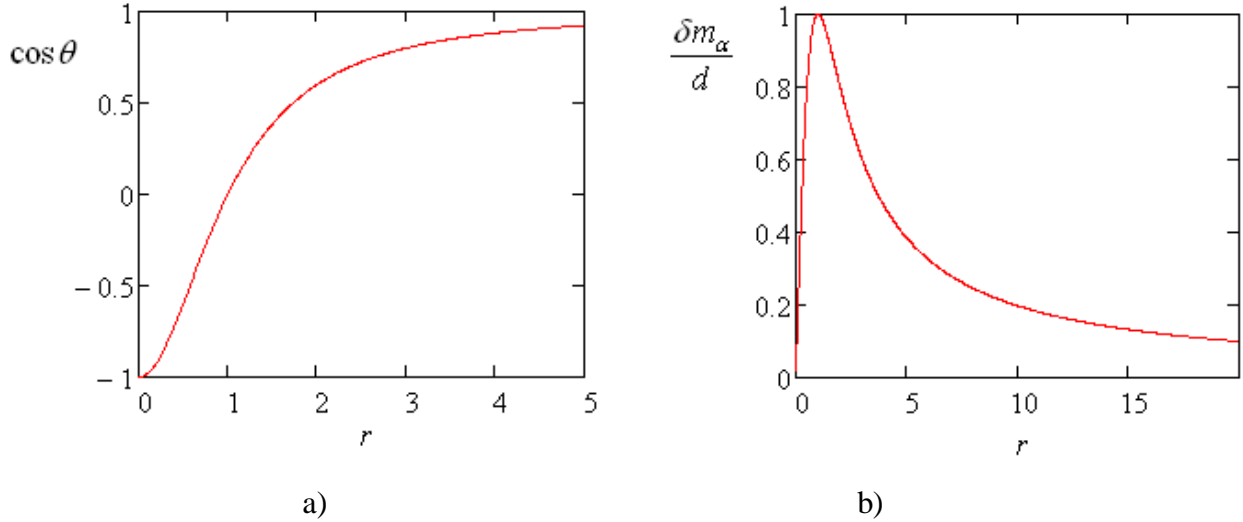

a)            b)

Fig. 1. a) $l_z = \cos\theta$ as a function of $r = \dfrac{|\vec{r}|}{R}$ at $Z = 0$ according to the formula (23); b) $\dfrac{\delta m_\alpha}{d}$ as a function of $r = \dfrac{|\vec{r}|}{R}$ at $Z = 0$ according to the formula (24).

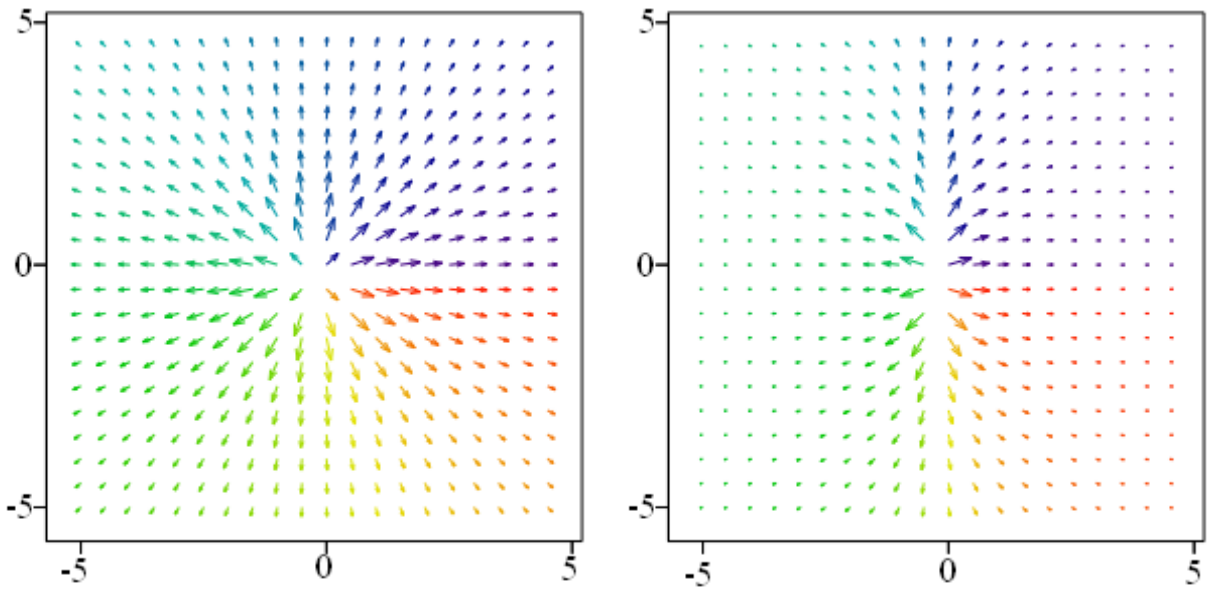

Fig. 2. a) Vector plot of projection of vector $\vec{l}$ on a plane $XOY$ at a) $v = 0$ according to the formula (23); b) $v = 0.95c$ according to the formula (23).

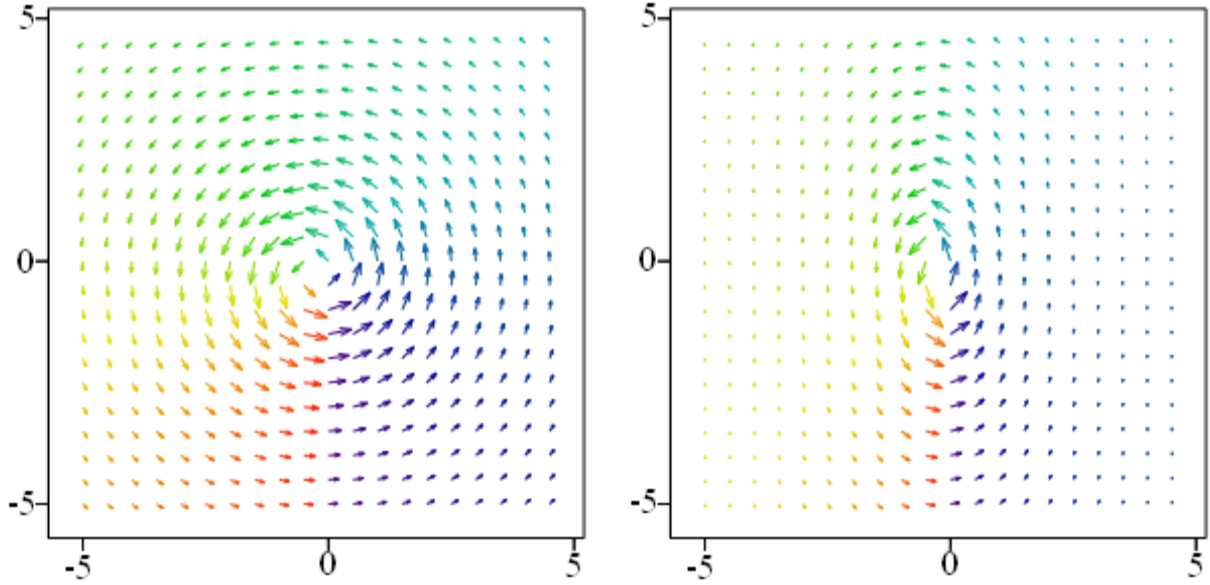

Fig. 3. a) Vector plot of projection of vector $\vec{m}$ on a plane *XOY* at a) $v=0$ according to the formula (24); b) $v=0.95c$ according to the formula (24).

Let us not that the equations (12) in the static case (i.e. at $\frac{\partial \theta}{\partial t}=0$ and $\frac{\partial \varphi}{\partial t}=0$) в coincides with the corresponding equations for polar and azimuth angles of magnetization vector in a ferromagnet with uniaxial magnetic anisotropy under exchange approximation (i.e. in the case when it is possible to neglect the magnetostatic energy of a ferromagnet). That is why the functional form of the static solutions of this work (12) is applicable in the case of static magnetization distributions of a ferromagnet with easy axis and easy plane magnetic anisotropy. The particular cases of the solutions (13), (15) in a ferromagnet include the well-known one dimensional solutions such as Bloch domain [38], Neel domain wall [39], Shirobokov domain structure [40], antiferromagnetic vortices [41, 42], two dimensional Belavin-Polyakov soliton [43], three dimensional Hodenkov soliton [44] and target type soliton [45].

The results of this paper can be used for further development of theory of the antiferromagnetic soliton physics [28, 46, 47] and theory of topological spin-hall [48] and spin torque [49, 50] effects at antiferromagnetic skyrmions and domain walls with skyrmion-like structure under "subcritical" and "supercritical" modes of movement.